\documentclass[aps,superscriptaddress,pra,preprint]{revtex4-1}
\usepackage{amsmath}
\usepackage{graphicx}
\usepackage{color}
\usepackage{amssymb}
\usepackage{gensymb}

\begin{document}

\title{The FLAME-slab method for electromagnetic wave scattering in aperiodic slabs}

\author{Shampy Mansha}

\address{Division of Physics and Applied Physics, School of
  Physical and Mathematical Sciences, Nanyang Technological
  University, Singapore 637371, Singapore}

\author{Igor Tsukerman}

\address{Department of Electrical and Computer Engineering, The
  University of Akron, Akron, OH 44325-3904, USA}

\author{Y.~D.~Chong}

\address{Division of Physics and Applied Physics, School of
  Physical and Mathematical Sciences, Nanyang Technological
  University, Singapore 637371, Singapore}

\address{Centre for Disruptive Photonic Technologies, \\ Nanyang
  Technological University, Singapore 637371,
  Singapore}

\email{yidong@ntu.edu.sg}

\date{\today}

\begin{abstract}
The proposed numerical method, ``FLAME-slab,'' solves electromagnetic
wave scattering problems for aperiodic slab structures by exploiting
short-range regularities in these structures.  The computational
procedure involves special difference schemes with high accuracy even
on coarse grids.  These schemes are based on Trefftz approximations,
utilizing functions that locally satisfy the governing differential
equations, as is done in the Flexible Local Approximation Method
(FLAME). Radiation boundary conditions are implemented via Fourier
expansions in the air surrounding the slab. When applied to ensembles
of slab structures with identical short-range features, such as
amorphous or quasicrystalline lattices, the method is significantly
more efficient, both in runtime and in memory consumption, than
traditional approaches. This efficiency is due to the fact that the
Trefftz functions need to be computed only once for the whole
ensemble.
\end{abstract}

\maketitle

\section{Introduction}

\subsection{Electromagnetic wave scattering in slab structures}
\label{sec:flame_intro}

\begin{figure}[b]
  \centering
  \includegraphics[width=10.9cm]{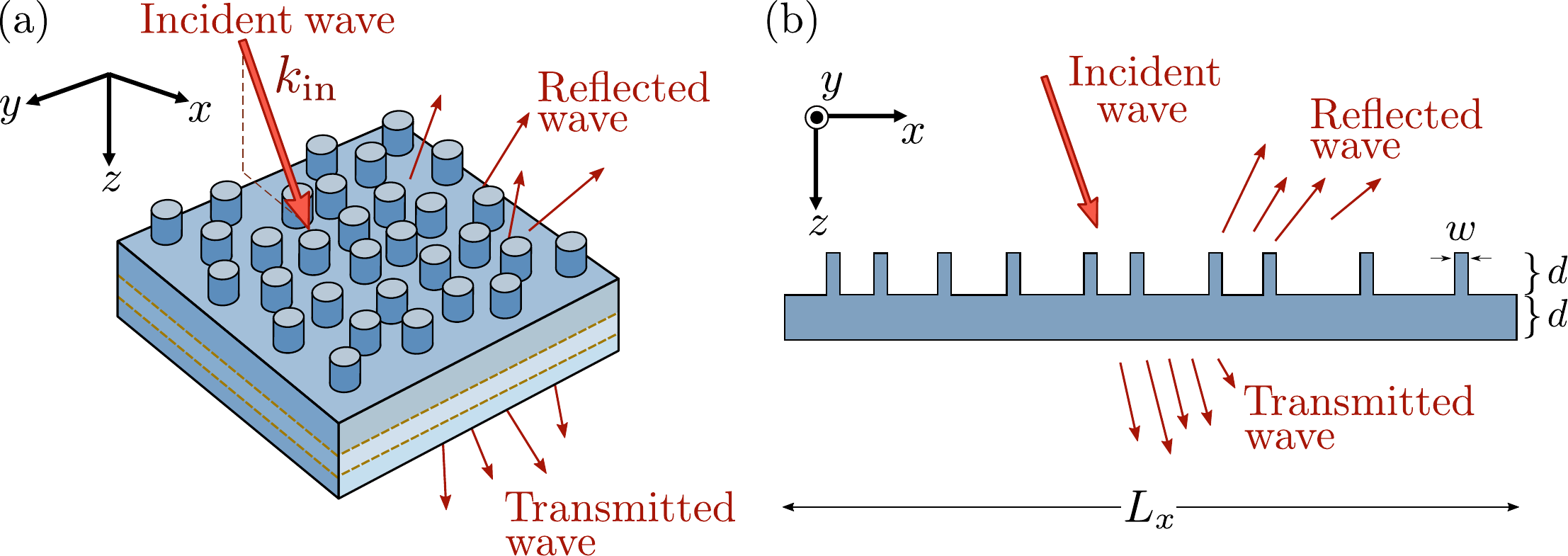}
  \caption{(a) Schematic of electromagnetic wave scattering from a 3D slab. The slab consists of one or more 
  	dielectric layers stacked in the $z$ direction, and is patterned (e.g., with pillars) along the $x$-$y$ plane.  
  	A plane wave is incident from the $-z$ side, with a wave vector $\mathbf{k}_{\mathrm{in}}$.  
  	(b) The 2D version of the problem, where the dielectric function does not vary with $y$, 
  	and the wave is incident along the $x$-$z$ plane. }
  \label{fig:schematic}
\end{figure}

Scattering of electromagnetic waves by a patterned photonic slab is
an important but challenging class of problems in computational
electromagnetics. As shown in Fig.~\ref{fig:schematic}, such a
structure consists of one or more layers of finite thickness, stacked
along the ``vertical'' axis ($z$).  Along the other two directions (in
the ``horizontal'' $x$-$y$ plane), the slab is effectively infinite in
extent, but is patterned with structural elements such as pillars or
holes, making it spatially non-uniform \cite{JDJ2011,Erchak2001}. Slab
structures form a basis for a wide range of important photonic
devices, including photonic crystal slab lasers
\cite{Painter1999,Noda2001,Imada2002,Ryu2002,Matsubara2008,chua2011}
and metasurfaces \cite{Yu2014,Chen2016}.

This problem is commonly tackled using standard finite element (FE)
or finite difference (FD) discretization, with absorbing conditions
or perfectly matched layers (PMLs) to simulate the infinite free-space
region outside the slab. There also exist two special computational
methods tailored to slabs with piecewise-uniform dielectric functions:
Rigorous Coupled Wave Analysis (RCWA) \cite{Moharam1995,Liu2012} and
the Vertical Mode Expansion Method (VMEM) \cite{Lu2014,Shi2016}. In
RCWA, 2D Fourier transforms are performed in each horizontal layer,
and the solutions are matched along the layer boundaries. In VMEM, a
mode expansion is performed on a set of vertical boundaries (e.g.,
cylinder or hole surfaces), and boundary integral equations are used
to match the expansions in the horizontal plane.  All of these methods
are fairly efficient for periodic structures, where the computational
domain can be reduced to a single horizontal lattice cell.  (VMEM can
also be applied to a standalone feature such as a single cylinder or
hole \cite{Lu2014}.)

The computational problem becomes substantially more difficult when
there is no horizontal periodicity.  In recent years, there has been
increasing research interest in photonic structures that are
deliberately patterned in an aperiodic manner, including random or
amorphous patterns
\cite{Ballato1999,Jin2001,Garcia2007,Florescu2009_2,Noh2011_1,
  Liang2013, Mansha2016}, quasicrystalline patterns
\cite{Feng2005,Steurer2007,Ricciardi2009,Yang2010,Vardeny2013,DalNegro2013},
and topological defects \cite{Liew2015,Knitter2015}.  Such structures
have a variety of interesting properties; for instance, amorphous
photonic structures have been found to exhibit highly isotropic
spectral gaps \cite{Jin2001}, while slabs patterned with topological
defects have been shown to support laser modes with inherent
power-flow vorticity \cite{Knitter2015}.  To model these various types
of aperiodic slab structures, it is generally necessary to define a
horizontal supercell, which in turn requires a large mesh (for FE and
FD methods) or a high-order series expansion (for RCWA and VMEM).
Such approaches fail to exploit a common feature in most of these
examples, which is the existence of short-range regularity.  Amorphous
and quasicrystalline patterns, for instance, are often generated by
taking one or more fixed elements---e.g., pillars or holes with a
single fixed radius---and assigning them to aperiodic positions
\cite{Jin2001,Feng2005}.

This paper explores a new computational method adapted to wave
scattering problems in slab structures where the horizontal patterning
has short-range regularity but no long-range periodicity, and hence no
simple unit cell.  We call this method ``FLAME-slab'' because its core
component is high-order finite difference (FD) schemes generated by
the Flexible Local Approximation MEthod (FLAME)
\cite{Tsukerman2005,Tsukerman2006}. FLAME replaces the local
polynomial expansions of classical FD with more general, and often
much more accurate, approximations by Trefftz functions, which by
definition satisfy the governing equation of the problem and the
applicable interface boundary conditions (b.c.)---in our case,
Maxwell's equations and their associated b.c.

Previously, Trefftz approximations have been successfully used in
classical variational contexts (FE and domain decomposition methods)
\cite{Jirousek1978,Herrera2000,Qin2000}, in Discontinuous Galerkin
methods \cite{Hiptmair2013,Kretzschmar2013,Egger2015,Egger2015a}, and
in
FLAME~\cite{Tsukerman2005,Tsukerman2006,Webb2009,Classen2010,Kretzschmar-Springer15,Tsukerman2016};
see also review \cite{Hiptmair-Springer16} by Hiptmair \textit{et
  al}.

In FLAME-slab, as explained in detail in the following sections, the
Trefftz basis functions are generated via small-scale simulations
performed by an auxiliary method such as RCWA. The key point is that
the bases and the corresponding difference schemes are purely local,
and can thus be computed on geometrically small but physically
representative segments of the overall structure; we shall call these
segments ``Trefftz cells''.  (Of course, a global FD scheme for the
whole structure must still be assembled and solved, but many standard
numerical linear algebra techniques can be brought to bear on that
part of the computational procedure.)  Assuming the slab has the
aforementioned short-range regularity, only a small number of basis
functions need to be computed.  Moreover, the basis functions can be
cached and re-used for different slab structures possessing the same
local parameters.

To illustrate the method, in Section~\ref{sec:slab} we formulate
FLAME-slab for a 2D structure with a large number of
randomly-positioned rectangular pillars, as depicted in
Fig.~\ref{fig:schematic}(b).  In this case, the Trefftz basis
functions can be generated from RCWA simulations of one pillar with a
surrounding section of the substrate.  Although this reference problem
is 2D, FLAME-slab is intended to be generalizable to fully 3D
geometries.  In Section~\ref{sec:FLAMEFFT_results}, we show that the
FLAME-slab results agree with RCWA calculations on the full structure.
When the FLAME-slab solver re-uses basis functions, its runtime is
around two orders of magnitude lower than RCWA alone.

We remark in passing that Trefftz basis functions can also be
generated using other numerical methods, such as FEM or classical FD,
as well as VMEM~\cite{Lu2014,Shi2016}.  We have opted for RCWA because
of its advantages: it does not require spatial discretization of the
vertical direction (so long as the slabs are piecewise-uniform),
handles the scattering boundary conditions exactly, without PML
approximations, and is available as free software \cite{Liu2012}.

\subsection{Formulation of the scattering problem}
\label{sec:formulation}

We consider a slab of a finite thickness $d_{\mathrm{slab}}$ in free space, extending over $0 \leq z \leq d_{\mathrm{slab}}$.  The slab has magnetic permeability $\mu = 1$ everywhere, and can be characterized by a (possibly complex) dielectric permittivity $\epsilon(\omega, \mathbf{r})$.  In this paper, the problem is formulated in the frequency domain, and dependence of parameters on the frequency $\omega$ will not henceforth be explicitly indicated.  No theoretical limitations on the position dependence of $\epsilon(\mathbf{r})$ are imposed, although in practice $\epsilon(\mathbf{r})$ is typically piecewise-constant, representing a modest number of different constituent materials in an actual device.

The present paper demonstrates a proof-of-concept implementation of
FLAME-slab for an effectively-2D geometry. As shown in
Fig.~\ref{fig:schematic}(b), the slab is assumed to be translationally
invariant along $y$.  The computational domain is assumed to be a
supercell, with quasi-periodic boundary conditions relating the fields
on the left ($x=0$) and right ($x=L_x$) sides. These conditions do not
necessarily reflect an actual periodicity of the sample, but are
imposed by necessity, to mimic the field behavior in an infinite (or very large) structure with as few numerical artifacts as possible.
 
The electromagnetic field in and around the slab is governed by Maxwell's equations,
\begin{equation}
  \nabla \times \mathbf{E} = i k_0 \mathbf{H}, \quad \nabla \times \mathbf{H} = -i k_0 \varepsilon \mathbf{E},
  \label{eq:Maxwell}
\end{equation}
where $k_0 = \omega / c$ is the wavenumber in free space.  The
$\exp(-i \omega t)$ phasor convention is adopted.  We consider the
$s$-mode, with a one-component electric field $\mathbf{E} = E\hat{y}$
and a two-component magnetic field $\mathbf{H}= H_x \,\hat{x} + H_z
\,\hat{z}$.  Further, we assume that the slab is illuminated by a
plane wave incident from $z < 0$:
\begin{equation}
  E_{\mathrm{in}} (\mathbf{r}) \,=\, \exp(i \mathbf{k} \cdot r),
  ~~~ k_x = k_0 \sin \theta_{\mathrm{in}}, ~~ k_z = k_0 \cos \theta_{\mathrm{in}},
  \label{eq:plane-wave}
\end{equation}
where $\theta_{\mathrm{in}}$ is the angle of incidence. The
quasi-periodic boundary conditions on the left and right sides of the
computational domain are
\begin{equation}
  E(L_x, z) \,=\, E(0, z) \exp(i k_x L_x), ~~~
  \mathbf{H}(L_x, z) \,=\, \mathbf{H}(0, z) \exp(i k_x L_x).
  \label{eq:bc-left-right}
\end{equation}
Finally, we introduce the usual splitting of the total fields into the incident and scattered parts. The formal definition of the scattered field is
\begin{equation}
  E_s (\mathbf{r}) = E(\mathbf{r}) - E_{\mathrm{in}} (\mathbf{r}).  
\end{equation}
$E_s$ is subject to the standard radiation boundary conditions ensuring that $E_s$ 
propagates away from the slab (i.e., toward $z \rightarrow +\infty$ for 
$z > d_{\mathrm{slab}}$ and toward $z \rightarrow -\infty$ for $z < 0$).

\subsection{Rigorous Coupled Wave Analysis (RCWA)}
\label{sec:RCWA}

In this section, we briefly review the Rigorous Coupled Wave Analysis
(RCWA) algorithm for solving the electromagnetic wave scattering
problem in slabs \cite{Moharam1995,Liu2012}.  Our implementation of
FLAME-slab uses the RCWA solver $S^4$ (the Stanford Stratified
Structure Solver) \cite{Liu2012} to calculate Trefftz basis functions,
and also to provide comparisons for the numerical results.

The slab (see Fig.~\ref{fig:schematic}) is divided into layers in the
vertical ($z$) direction.  It is assumed that the dielectric function
$\varepsilon(x,y,z)$ is periodic in the $x$-$y$ plane (if necessary,
by defining a supercell), and piecewise-uniform in the $z$ direction.
We let $\mathbf{k}_\parallel$ denote the projection of the incident
wave-vector, $\mathbf{k}_{\mathrm{in}}$, onto the plane. In-plane
Fourier decomposition of the electromagnetic field in each layer
\cite{Li1997} is then carried out.  The electric field is expanded in
the form
\begin{equation}
  \mathbf{E}(\mathbf{r}_\parallel,z) = \sum_{\mathbf{G}} {\widetilde{\mathbf{E}}}_{\mathbf{G}} \, \exp\big[iq_z z\big]  \;
  \exp\big[i(\mathbf{k}_\parallel+\mathbf{G})\cdot\mathbf{r}_\parallel\big],
  \label{eq:Hexpansion}
\end{equation}
where $\mathbf{r}_\parallel = (x,y)$ denotes in-plane coordinates,
$\{\mathbf{G}\}$ is the set of reciprocal lattice vectors,
$\widetilde{\mathbf{E}}_{\mathbf{G}}$ is a Fourier expansion
coefficient, and $q_z$ is the Bloch wavenumber in the $z$ direction.
The magnetic field is expanded similarly.  The in-plane Fourier series
are truncated by specifying an upper bound for $|\mathbf{G}|$.  The
number of terms retained is denoted by $N_G$, so that $N_G \sim
|\mathbf{G}|^2$ for a 3D problem (with 2D periodicity), and $N_G \sim
|\mathbf{G}|$ for 2D (with 1D periodicity).

The Fourier harmonics are then matched across adjacent layers via the
transfer matrix method \cite{Whittaker1999}.  After repeatedly
applying the procedure to each layer, one obtains a transfer matrix
relation between the fields at the outermost layers of the slab, and
hence the reflection and transmission coefficients.  From these, all
other quantities of interest can be retrieved, including the fields
within the slab as well as the far field.  For details, see
\cite{Moharam1995,Liu2012}.

For our present purposes, the key point to note is that the runtime of the RCWA calculation depends principally on the integer $N_G$, the number of reciprocal lattice vectors included in the in-plane Fourier expansion \eqref{eq:Hexpansion}.  The method involves running an eigensolver on a full matrix with $\mathcal{O}(N_G)$ rows/columns.  Hence, the scalings of the runtime $T_{\mathrm{RCWA}}$ and the memory usage $M_{\mathrm{RCWA}}$ are
\begin{equation}
  T_{\mathrm{RCWA}} \sim \mathcal{O}(N_G^3), \quad M_{\mathrm{RCWA}} \sim \mathcal{O}(N_G^2).
  \label{eq:RCWA_runtime}
\end{equation}
We have verified that the performance of the $S^4$ solver
\cite{Liu2012} is consistent with these scalings.

\subsection{FLAME}
\label{sec:flame1d}

In this section, we give a brief general description of FLAME
\cite{Tsukerman2005,Tsukerman2006}, and explain how it can be
specialized to the slab problem under consideration.  This exposition
follows previous publications on this subject, especially
\cite{Kretzschmar-Springer15,Tsukerman2016}.

Consider a differential equation of the form
\begin{equation}
  \mathcal{D}(\mathbf{r}) \;u(\mathbf{r}) = 0,
  \label{eq:fdproblem}
\end{equation}
where $\mathcal{D}(\mathbf{r})$ is a differential operator, and
$u(\mathbf{r})$ is either a scalar or vector field.  Mathematically,
Eq.~\eqref{eq:fdproblem} should be understood in weak form: in
physical terms, it includes proper interface boundary conditions in
addition to the differential equation itself.

The general setup for FLAME has two ingredients: (i) a set of $n$
local Trefftz functions solving Eq.~\eqref{eq:fdproblem} within small
local subdomains, or ``patches'', of the computational domain (each
patch contains a grid ``molecule''---a set of geometric entities such
as grid nodes---on which the difference scheme is to be formed), and
(ii) a set of $m$ degrees of freedom (DoF), also defined for each
patch.  These DoF are, by definition, linear functionals, $l_\beta(u)$
($\beta = 1,2,\dots, m$), each mapping the field $u$ to a number.  The
term ``DoF'' is used here in its mathematical sense, rather than that
of kinematics or statistical physics; in the simplest case, if $u$ is
a scalar field, we might define $l_\beta(u) \equiv
u(\mathbf{r}_\beta)$, where $\mathbf{r}_1,\dots,\mathbf{r}_m$ are a
set of grid nodes.  Other examples of DoF include the circulation of a
field over a given path, the flux over a given surface, the nodal
value of any partial derivative of a field component, and so on.

Within each patch, the solution $u$ is approximated locally using a
linear combination of Trefftz functions $\varphi_\alpha$ ($\alpha =
1,2,\dots, n$) :
\begin{equation}\label{eqn:uh-eq-c-psi}
u(\mathbf{r}) \,\approx\, u_h(\mathbf{r}) \equiv \sum_\alpha c_\alpha \varphi_\alpha (\mathbf{r}) \,=\,
\mathbf{c}^T \boldsymbol{\varphi}(\mathbf{r}),
\end{equation}
where $\mathbf{c} \in \mathbb{C}^n$ is a coefficient vector and
$\boldsymbol{\varphi}$ is a vector of basis functions (both generally
complex).  For each patch (i.e. for each grid ``molecule'') we seek an
FD equation of the form
\begin{equation}\label{eqn:s-beta-l-beta-eq-0}
   \sum_{\beta=1}^m s_\beta l_\beta (u) = 0,
\end{equation}
where $\mathbf{s} = (s_1, s_2,\dots, s_m)^T$ is a vector of complex
coefficients (a ``scheme'') to be determined.  In the simplest version
of FLAME, the scheme is required to be exact for any linear
combination \eqref{eqn:uh-eq-c-psi} of basis functions (Least-squares schemes can also be useful
\cite{Boag94,Tsukerman-JCP10}).
After straightforward algebra, it can be shown that \cite{Tsukerman2005,Tsukerman2006}
\begin{equation}\label{eqn:s-in-null-Nt}
  \mathbf{s} \in \mathrm{Null}(N^T), \;\;\;\mathrm{where}\;\;\;  N^T_{\alpha \beta} = l_\beta (\varphi_\alpha).
\end{equation}
The construction of the scheme is
purely local, which is consistent with the local nature of the
differential operator $\mathcal{D}$ in Eq.~\eqref{eq:fdproblem};
therefore the choices of Trefftz bases and DoF in different patches
are independent of one another.  Specifically, in this paper, the
patches centered near the middle of the slab use different DoF from
those in the boundary patches. 

In Section~\ref{sec:slab}, we consider a 2D geometry with rectangular
pillars distributed on a slab substrate, shown in
Fig.~\ref{fig:schematic}(b).  We will define three types of patches;
one has DoF corresponding to nodal values of the $E$-field, while the
other two types have DoF consisting of nodal values of the $E$-field,
plus tangential $\mathbf{H}$ field components at the boundary nodes.
The assignment of these DoF is further elaborated on in
Section~\ref{sec:layout}.

Each Trefftz basis only needs to be \textit{local}, i.e.~valid over a
given patch.  Therefore, it can be generated by solving a set of
relatively small auxiliary problems.  For instance, for the slab
considered in Section~\ref{sec:slab}, the local problem can be
geometrically limited to ``Trefftz cells'' consisting of a single
pillar with a segment of the surrounding substrate.  We obtain basis
functions by solving the wave scattering problem in the Trefftz cells,
with different illumination conditions.  A large part of the
computational savings comes from the fact that the local problem is
much smaller in size than the global one.  This procedure is described
in detail in Section~\ref{sec:flame_generation}.  In this paper, we
use RCWA to solve the local problem, but in principle any suitable
method could be used.

Once a Trefftz basis and DoF have been determined for any given patch,
Eq.~\eqref{eqn:s-in-null-Nt} yields the local difference scheme
immediately.  In our setup, the scheme has nine terms in each patch,
both inside the slab and at its boundary.  (It is tacitly assumed that
the null space of $N^T$ is one-dimensional.)  The FLAME block of the
global FD matrix is then assembled from the local schemes in a
standard way, as is done in all FD algorithms; see
Section~\ref{sec:assembly}.

\section{Finite-difference schemes for slab geometry}
\label{sec:slab}

\subsection{Discretization and Trefftz patches}
\label{sec:layout}

\begin{figure}
\centering
\includegraphics[width=10cm]{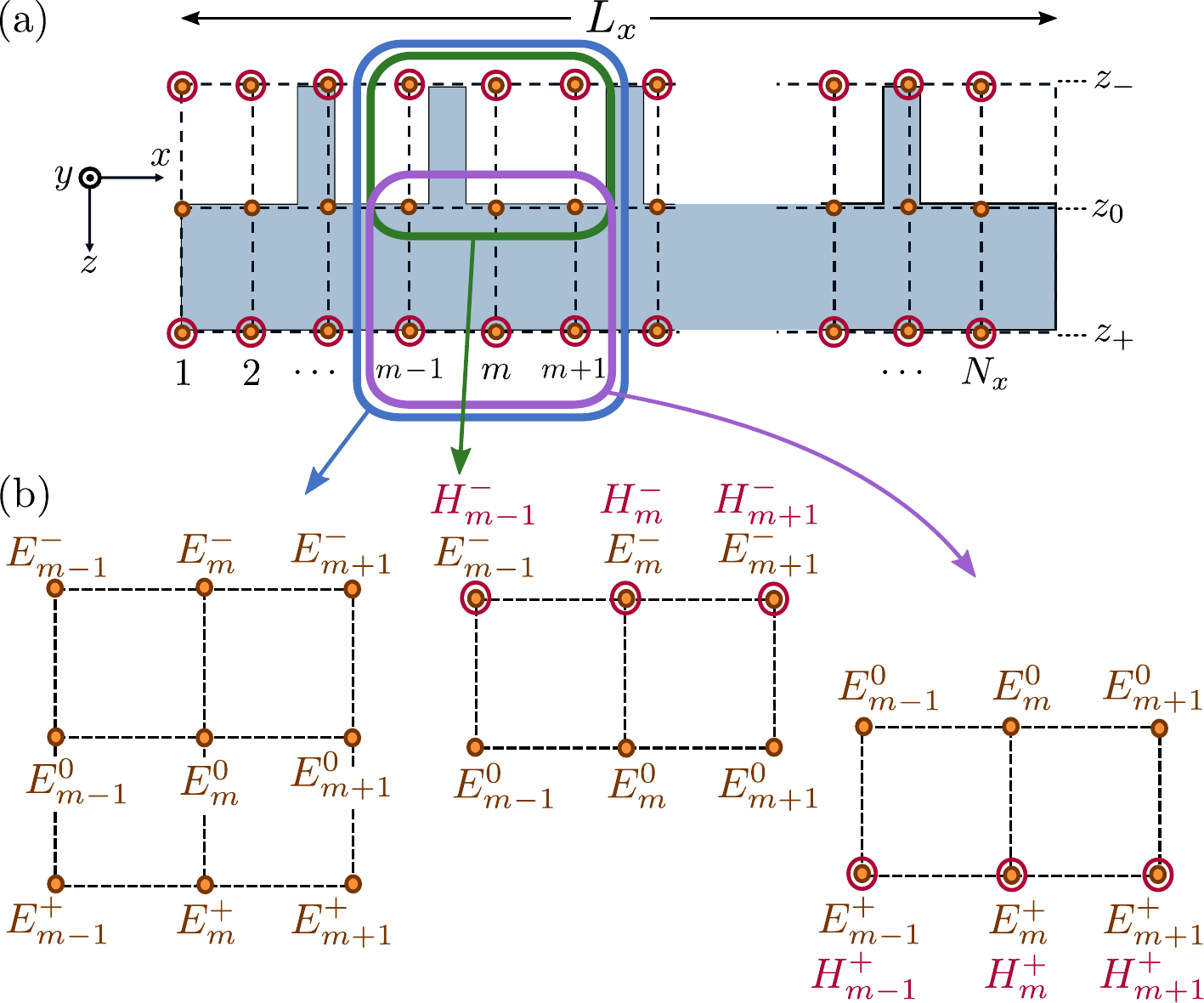}
\caption{(a) A coarse $N_x \times N_z$ grid for the slab structure,
  where $N_z = 3$.  (b) Assignment of DoF for patches centered on
  column $m$.  Left: patch centered at $z^0$, whose DoF are the
  $y$-components of the electric field in the nine nodes.  Center and
  right: boundary patches, whose DoF are the electric fields at the
  six nodes and the $x$-component of the magnetic field at the
  boundary nodes.}
\label{fig:model}
\end{figure}

Our FLAME-slab implementation applies to the 2D slab geometry
described in Section~\ref{sec:formulation} and depicted in
Fig.~\ref{fig:schematic}(b).  We adopt the
convenient discretization shown in Fig.~\ref{fig:model} (a).  To
demonstrate that sufficient accuracy can be achieved in FLAME-slab, we
have deliberately chosen only three grid layers ($N_z = 3$) across the
thickness of the slab, including the two outer layers.  These outer
layers of nodes lie in the air regions slightly above/below
the structure.  In the $x$ direction, the 2D computational domain (or
``supercell'') of total length $L_x$ contains $N_x$ grid lines, which
are equally-spaced for simplicity. Importantly, the grid need not
resolve the fine geometric features of the structure, as the relevant
information will be contained in the Trefftz bases.  In particular,
even though we only use three layers of grid points, the slab could
consist of multiple layers of dielectric material, and the grid lines
need not coincide with any material interfaces.

Figure~\ref{fig:model}(b) shows how the DoF are assigned in each patch.
There are three types of patches.  The first type contains $3 \times 3
= 9$ nodes on adjacent grid lines, and we take its DoF to be the nine
nodal values of the \textit{total} $E$ field, $E^{-,0,+}_{m-1,m,m+1}$.
(The splitting of the field into ``incident'' and ``scattered'' parts
is only needed later, to impose the radiation boundary conditions; see
Section~\ref{sec:scattering}.)  The second and third types of patch
contain $2\times3 = 6$ nodes on the upper or lower boundary; for each,
six DoF are the respective values of the $E$ field, and
three more DoF are the values of the magnetic field component $H_x$,
tangential to the slab, at the boundary nodes $m-1, m, m+1$.

The total number of unknowns in the system of equations to be solved
is $N_x (N_z+2)$; in the present setup, with $N_z = 3$, this number is
$5N_x$. Of these unknowns, $N_x \cdot N_z$ are the values of the $E$
field at all nodes, and the remaining $2N_x$ are the values of $H_x$
at the boundary nodes.  Correspondingly, $N_x \cdot N_z$
\textit{equations} of the system are the Trefftz-FLAME difference
schemes described above. The remaining $2N_x$ equations come from the
radiation boundary conditions, as explained in
Section~\ref{sec:scattering}.

In matrix form, the FD-FLAME equations can be written as
\begin{equation}
\mathbf{A}_{\mathrm{FLAME}} \boldsymbol{\psi} = 0,
\label{eq:Apsi_total}
\end{equation}
where $\mathbf{A}_{\mathrm{FLAME}}$ is a rectangular $N_x N_z\times N_x(N_z+2)$ matrix 
and $\boldsymbol{\psi}$ is a vector containing all $N_x (N_z + 2)$ DoF (nodal values of the fields).

\subsection{Generation of Trefftz basis functions}
\label{sec:flame_generation}

\begin{figure}[b]
\centering
\includegraphics[width=8cm]{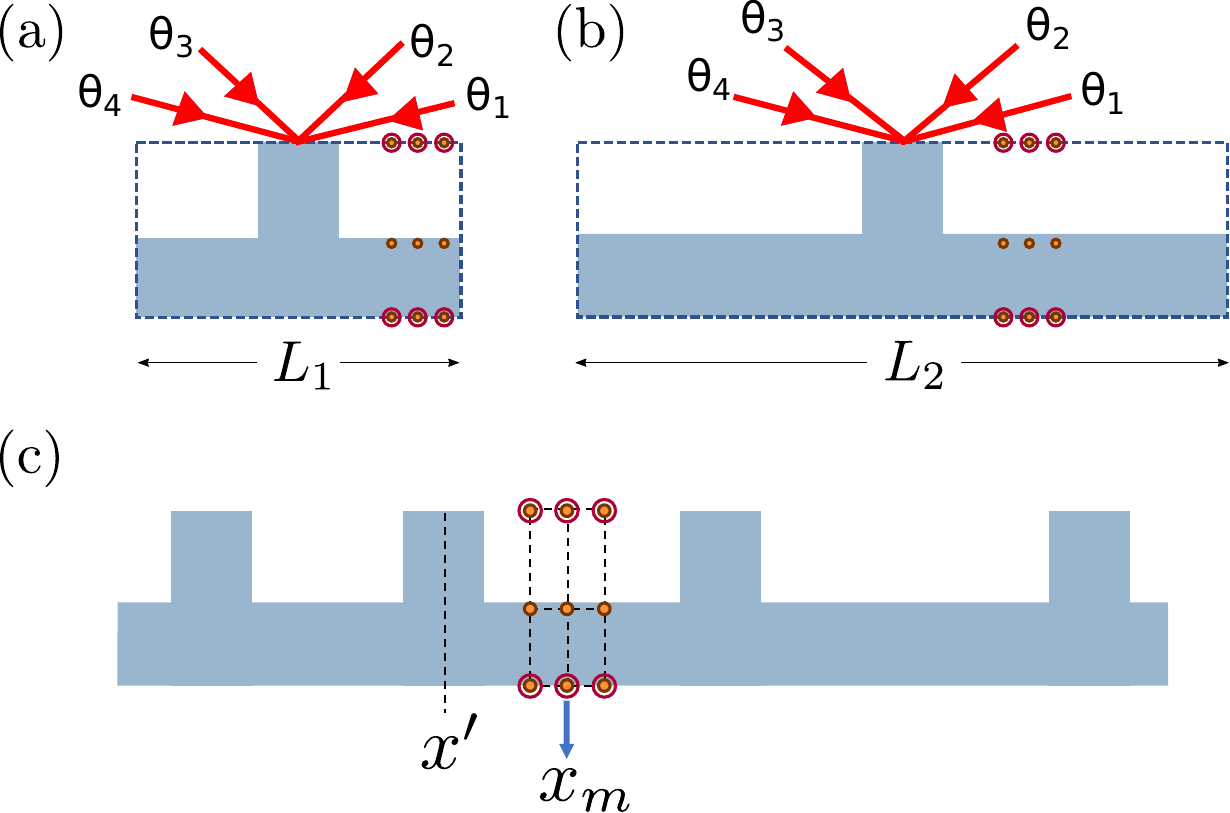}
\caption{Procedure for generating Trefftz basis functions.
    (a)--(b) Schematic of ``Trefftz cells'' defined as unit cells of
    periodic slabs with one pillar per period.  Two periods are used,
    $L_1$ and $L_2$; for each, the RCWA solver produces four Trefftz
    functions corresponding to angles of illumination $\theta_{1},
    \dots, \theta_{4}$.  (c) For a given patch centered at grid
    coordinate $x_m$, one uses the eight functions precomputed in (a,
    b) according to the position $x'$ of the nearest pillar.}
\label{fig:lattice_period}
\end{figure}

As already noted, the computational savings in FLAME-slab come from
the fact that Trefftz functions in FD-FLAME only need to be local, and
can thus be computed over a relatively small representative segment of
the structure. (In practice, they can be \textit{pre}-computed and
stored.)  For the present 2D slab geometry, we define a ``Trefftz
cell'' consisting of a single pillar, with some surrounding substrate
of a length $L$.  By solving Maxwell's equations in the Trefftz cell
for different illumination and boundary conditions, we produce the
desired Trefftz basis set.

A natural set of illumination conditions comes from plane waves
impinging on the segment of the slab at different angles, as
schematically shown in Fig.~\ref{fig:lattice_period}.  The scattered
wave is subject to the standard radiation conditions above and below
the slab. One also needs to impose some boundary conditions on the
left and right sides of the segment. Extensive computational evidence
in photonics, molecular dynamics, and other areas---where very large
structures are modeled via a finite representative sample---has shown
that the least intrusive conditions are (quasi-)periodic ones. In our
case, such conditions read
\begin{align}
  \begin{aligned}
    E(x = \frac{L}{2}, \, z) &= \exp(i k_0 L \cos \theta_i) \, E(x = -\frac{L}{2}, \, z) \\
    H_x(x = \frac{L}{2}, \, z) &= \exp(i k_0 L \cos \theta_i) \, H_x(x = -\frac{L}{2}, \, z)
    \label{eq:bc-segment}
  \end{aligned}
\end{align}
where $x$ is the ``horizontal'' coordinate relative to the pillar, and
$\theta_i$ is the angle of illumination. Due to these quasi-periodic
conditions, the segment of length $L$ can be viewed as the unit cell
of a fictitious photonic crystal. The fact that the actual amorphous
structure under investigation does not have this periodicity is
unimportant because the Trefftz basis approximates the solution only
locally, over a small patch, and its long-range behavior is
inconsequential.  The fictitious period $L$ is much smaller than the
computational domain, but its dielectric function matches the actual
structure in the vicinity of a given pillar.

In our examples, nine-point FD stencils are used, and eight Trefftz
basis functions are needed to produce a valid FLAME scheme. There are
several ways to accomplish this.  We settled on the use of two
different lengths $L$ \eqref{eq:bc-segment} for the local lattice,
$L_1$ and $L_2$, and four angles of illumination $\{\theta_{1}, \dots,
\theta_{4}\}$ for each of these lengths.  The $2 \times 4 = 8$
solutions (Trefftz functions) are precomputed with the RCWA solver
$S^4$ \cite{Liu2012}, and stored.  Using these, we can apply the FLAME
equation \eqref{eqn:s-in-null-Nt} to all grid stencils, as described
in Section~\ref{sec:layout}, thereby populating the $3N_x$ rows of the
$\mathbf{A}_{\mathrm{FLAME}}$ matrix \eqref{eq:Apsi_total}.

%% For instance, Eq.~\eqref{eq:stencil1} involves only the 9 $E$-field variables, denoted by $E_{m+n}^\mu$ where $n \in \{-1,0,1\}$ and $\mu \in \{+,0,-\}$.  We denote the $E$-field variables in the 8 independent RCWA solutions ($p=1,\dots,8$) by $\mathcal{E}_{m+n}^{\mu,p}$.  These are used as a basis for the actual $E$-field within the patch:
%% \begin{equation}
%%   E_{m+n}^\mu = \sum_{p=1}^8 \alpha_p \, \mathcal{E}_{m+n}^{\mu, p} = \boldsymbol{N} \boldsymbol{\alpha}.
%% \end{equation}
%% Here, $\boldsymbol{N}$ is a $9\times 8$ matrix containing the RCWA-derived solutions $\mathcal{E}_{m+n}^{\mu,p}$ (the row index is an aggregate of $\mu$ and $n$, while the column index is $p$).  We compute $\textrm{null}(\boldsymbol{N}^T)$, and the result is a set of stencil coefficients $\{s^{00}_{n}, s^{01}_{m+n}, s^{02}_{m+n}\}$ for Eq.~\eqref{eq:stencil1}.  This process is repeated for \eqref{eq:stencil2} and \eqref{eq:stencil3}.

We consider structures with short-range regularity, in the sense that
all the pillars are geometrically identical; only their placement is
random.  Many photonic structures of interest, including amorphous and
quasicrystalline structures
\cite{Ballato1999,Jin2001,Garcia2007,Florescu2009_2,Noh2011_1,Liang2013,Mansha2016,Feng2005,Steurer2007,Ricciardi2009,Yang2010,Vardeny2013,DalNegro2013,Liew2015,Knitter2015},
have similar forms of short-range regularity.  This feature is
exploited here to minimize the number of Trefftz functions that need
to be pre-computed; in our examples, this number is just eight for
fixed-frequency simulations.  Each solution is calculated with a high
spatial resolution, cached, and later retrieved during the actual
construction of the Trefftz-FLAME schemes.  This procedure is even
more appealing for the simulation of ensembles of multiple independent
realizations of the slab geometry (e.g., different samples of an
amorphous lattice with the same elementary structural parameters),
which is an important practical case. Then the cache can be re-used in
all of the calculations.  Note also that for certain quasicrystalline
designs, it may be advantageous to choose Trefftz cells with more than
one pillar; for example, for Fibonacci quasicrystals
\cite{Vardeny2013,DalNegro2013}, we might use Trefftz cells
corresponding to two different two-pillar segments.

\subsection{Radiation conditions}
\label{sec:scattering}

We apply the standard splitting of the electromagnetic field into incident and scattered parts;
this is needed because radiation conditions apply to the scattered component:
\begin{align}
  \begin{aligned}
    \mathbf{E} &= \mathbf{E}_{\mathrm{in}} + \mathbf{E}_{s}\\
    \mathbf{H} &= \mathbf{H}_{\mathrm{in}} + \mathbf{H}_{s}.
  \end{aligned}
  \label{eq:scattering_decomposition}
\end{align}
(These equations can be viewed as a formal definition of the scattered field.)
Applying this decomposition to Eq.~\eqref{eq:Apsi_total} yields
\begin{equation}
  \mathbf{A}_{\mathrm{FLAME}} \boldsymbol{\psi}_s = 
  - \mathbf{A}_{\mathrm{FLAME}} \boldsymbol{\psi}_{\mathrm{in}},
  \label{eq:Aflame_patch_definition}
\end{equation}
where $\boldsymbol{\psi}_{\mathrm{in}}$ and $\boldsymbol{\psi}_s$ are the decompositions of the vector of field components into incident and scattered parts.

Radiation conditions can be imposed via a Fourier decomposition in the
empty semi-infinite strips above and below the slab. This is facilitated
by the fact that the outer grid layers have been placed inside the free-space regions
rather than in the slab itself (see Section~\ref{sec:layout}). Maxwell's equations \eqref{eq:Maxwell}
relate $H_s(x,z)$ and $E_s(x,z)$ as
\begin{equation}
  H_s(x,z) = \frac{i}{\omega}\frac{\partial E_s}{\partial z}.
  \label{eq:HE_relation}
\end{equation}
where $H_s$ refers to the tangential ($x$) component of the magnetic
field, while $E_s$ is the electric field of the $s$-mode under
consideration.  For simplicity of presentation, we will describe the
case of normal incidence, $\theta_{\mathrm{in}} = 0$; then the
quasi-periodic boundary conditions turn into periodic ones.  The
Fourier expansion of the scattered field in the free space regions is
\begin{equation}
    E_s(x,z) = \sum_{n} c_n \, \exp\big[i(k_{nz}z+k_{nx}x)\big],
    \label{eq:Efield_expand}
\end{equation}
where $n$ runs over integer values and the horizontal wavenumbers
$k_{nx} = 2\pi n /L_x$.  

Combining Eqs.~\eqref{eq:HE_relation} and Eq.~\eqref{eq:Efield_expand} gives
\begin{equation}
  H_s(x,z) = -\frac{1}{\omega}\sum_{n} c_{n} \, k_{nz} \, \exp\left[i(k_{nx}x + k_{nz}z)\right].
  \label{eq:Hfield_expand}
\end{equation} 
From the dispersion relation in free space, $k_{nx}$ and $k_{nz}$ are related by
\begin{equation}
  k_{nz} = \pm \sqrt{k_0^2 - k^{2}_{nx}}.
  \label{eq:knz}
\end{equation}
The choice of the $\pm$ sign depends on which semi-infinite strip the
expression applies to.  Referring to the layout of
Fig.~\ref{fig:model}, the plus sign applies for $z = z_+$, and minus
for $z = z_-$.  These choices ensure that the scattered waves are
\textit{outgoing} from the slab.

We can now evaluate \eqref{eq:Efield_expand} and
\eqref{eq:Hfield_expand} along each outer layer of grid points, to
produce relations between the electric and magnetic field components:
\begin{align}
  \begin{aligned}
    H_s(x_m,z_\pm) &= 
    -\frac{1}{\omega} e^{-im\pi\left(1-N_x^{-1}\right)} \,
    \mathrm{FT}_m\Bigg\{k_{nz} \; \mathrm{FT}_n\Big\{E_s(x_{m'},z_\pm) e^{im'\pi\left(1-N_x^{-1}\right)} \Big\} \Bigg\}.
  \end{aligned}
  \label{eq:EH_ft_result}
\end{align}
Here, $k_{nz}$ uses the $\pm$ sign in Eq.~\eqref{eq:knz} corresponding
to $z_\pm$.  The notation $\mathrm{FT}_m\{f_n\}$ denotes the $m$-th
component of the discrete Fourier transform (DFT) on a vector with
components $\{f_n\}$.  The $e^{\pm i\pi(\cdots)}$ factors are used to
offset the DFT and center the spectrum at $k_{nx} = 0$, since we
follow the DFT convention
\begin{equation}
  \mathrm{FT}_m\Big\{f_n\Big\} = \sum_{n=0}^{N-1} e^{-2\pi i mn/N}\, f_n.
\end{equation}

\subsection{Assembling the finite-difference system}
\label{sec:assembly}

Equation~\eqref{eq:EH_ft_result} contributes $N_x$ equations for each outer
layer.  Combining this with Eq.~\eqref{eq:Aflame_patch_definition}
yields
\begin{equation}
  \begin{pmatrix} \mathbf{A}_{\mathrm{FLAME}} \\ \mathbf{A}_{\mathrm{BC}} \end{pmatrix} \; \boldsymbol{\psi}_s
  = \begin{pmatrix} -\mathbf{A}_{\mathrm{FLAME}} \boldsymbol{\psi}_{\mathrm{in}} \\ \boldsymbol{0}
  \end{pmatrix}.
  \label{eq:Adef}
\end{equation}
Here, $\mathbf{A}_{\mathrm{FLAME}}$ is a $N_x N_z \times N_x (N_z+2)$
sparse sub-matrix (in our numerical examples, $3N_x \times 5N_x$)
representing the FD-FLAME equations, and $\mathbf{A}_{\mathrm{BC}}$ is
a $2N_x \times N_x (N_z+2)$ (in our examples, $2N_x \times 5N_x$) full
sub-matrix representing the boundary conditions
\eqref{eq:EH_ft_result}. The vector $\boldsymbol{\psi}_{\mathrm{in}}$
contains the field components of the incident wave, evaluated at the
grid points.

\begin{figure}[b]
  \centering
  \includegraphics[width=9.5cm]{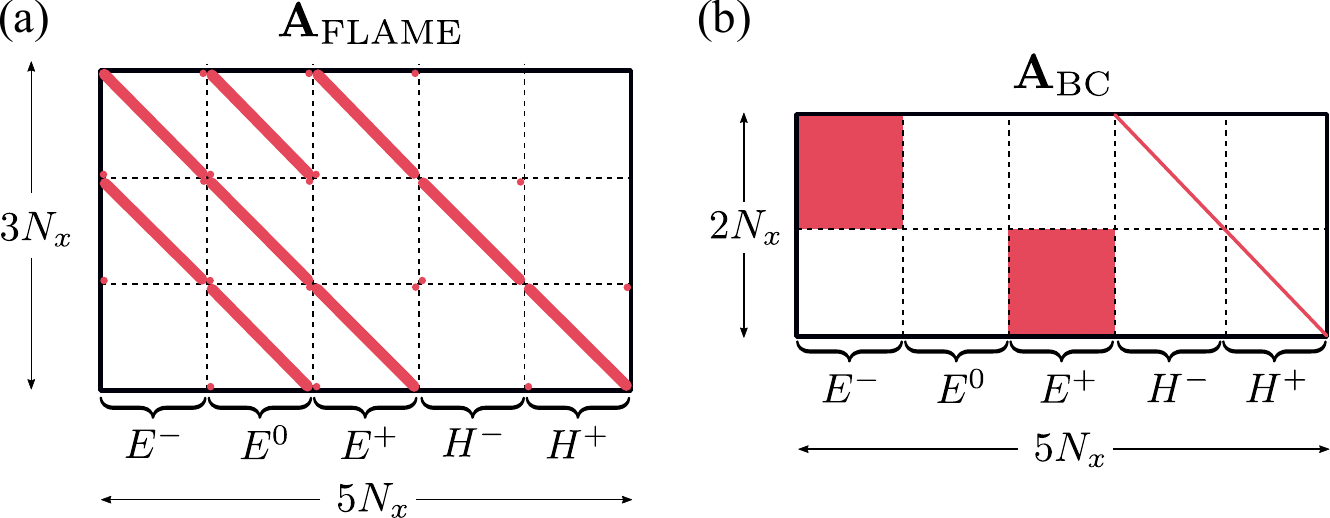}
  \caption{Structure of the matrix $\mathbf{A}$ for $N_z = 3$.
    Non-zero entries are indicated in red.}
  \label{fig:sparse_matrix}
\end{figure}

Figure~\ref{fig:sparse_matrix} illustrates the structure of the
sub-matrices $\mathbf{A}_\mathrm{FLAME}$ and $\mathbf{A}_\mathrm{BC}$,
for the test case where $N_z = 3$.  The sub-matrix
$\mathbf{A}_\mathrm{FLAME}$ contains nine nonzero entries per row,
based on the DoF assignment described in Section~\ref{sec:layout}.
The sub-matrix $\mathbf{A}_\mathrm{BC}$ has two full $N_x\times N_x$
subblocks, consisting of the Fourier components in
Eq.~\eqref{eq:EH_ft_result}.

Since $\mathbf{A}_{\mathrm{BC}}$ is full, its explicit calculation and
direct solution of Eq.~\eqref{eq:Adef} will require an overall runtime
of $\mathcal{O}(N_x^3)$.  A speedup can be achieved if, instead, an
iterative solver such as GMRES \cite{gmres} is applied to
Eq.~\eqref{eq:Adef}. Each iteration involves a matrix-vector product
which could be computed on the fly with Fast Fourier Transforms for
the dense blocks. The computational cost will be
$\mathcal{O}(N_{\mathrm{iter}} N_x^2 \log N_x )$, where
$N_{\mathrm{iter}}$ is the number of iterations.  With a right choice
of preconditioners, this could scale better than $\mathcal{O}(N_x^3)$.

\section{Results}
\label{sec:FLAMEFFT_results}

\subsection{Test problems}
\label{sec:parameters}

Our main test case is depicted in Fig.~\ref{fig:schematic}(b).  It
consists of 10 dielectric pillars placed randomly on a flat dielectric
substrate. Except where stated otherwise, we assume a normally
incident input wave whose frequency is $f = 0.25$.  We adopt a system
of units \cite{Liu2012} where the speed of light ($c$), vacuum
permeability ($\mu_0$) and vacuum permittivity ($\varepsilon_0$) are
all normalized to unity; this implies that $f=1/\lambda$, where
$\lambda$ is the free-space wavelength, i.e. the frequency has units
of inverse length. The normalized pillar height and substrate
thickness are both $d = 0.25$, and the pillar width is $w = 0.2$.  The
pillars and substrate have the same dielectric constant $\varepsilon =
12$, and are surrounded by air. The pillars are placed at aperiodic
positions on the substrate. The entire computational domain has length
$L_x = 14$ in the $x$ direction.

The horizontal grid layers are placed at $z_+ = 0$, $z_0 = 0.26$, and
$z_- = 0.5$.  For the Trefftz cells
(Section~\ref{sec:flame_generation}), we take $L_1 = 1.73$ and $L_2 =
4.12$, along with the incidence angles $\theta_1 = 36\degree$,
$\theta_2 = 72\degree$, $\theta_3 = 108\degree$, and $\theta_4 =
144\degree$, measured anti-clockwise from the $+x$ axis.

The accuracy of the FLAME-slab method depends on three adjustable
parameters, which we denote by $N_x$, $N_T$, and $N_G$.  The first,
$N_x$, is the number of nodes in the $x$ direction (see
Section~\ref{sec:layout}).  To make this physically meaningful, we
express it as the ratio $N_x w/L_x$, the number of nodes per pillar
width.  The second parameter, $N_T$, refers to the number of nodes in
the $x$ direction for the Trefftz cells. We normalize this parameter
as the ratio $N_Tw/L_1$, the number of nodes per pillar width for the
smaller of the two Trefftz cells.  The third and final parameter,
$N_G$, controls the accuracy of the RCWA method.  As described in
Section~\ref{sec:RCWA}, this corresponds to the number of reciprocal
lattice vectors included in the horizontal Fourier expansion.

\subsection{Electromagnetic field and reflectance calculations}
\label{sec:fields}

As described in Section~\ref{sec:layout}, each FLAME-slab calculation yields the scattered electric field $E_{y,s}$ along the three layers of grid points, denoted by $\{E_s^+, E_s^0, E_s^-\}$, as well as the scattered magnetic field $H_{x,s}$ along the two outer layers, denoted by $\{H_s^+, H_s^-\}$.

From these field components, we can extract two other quantities that
are of key interest in electromagnetic wave scattering problems: the
reflectance $R$ and transmittance $T$.  For $s$-polarized fields, the
$z$-component of the time-averaged Poynting vector is
\begin{equation}
  \langle S_z(x,z) \rangle = \frac{1}{2} \mathrm{Re} \Big[E(x,z)\, H^{*}(x,z)\Big],
\end{equation}
which involves precisely the field variables produced by FLAME-slab.
To obtain $R$, we average $S_z$ over the $N_x$ grid points along $z =
z_-$, using only the scattered fields.  The calculation of $T$ uses
the \textit{total} fields along $z = z_+$, the side of the slab
opposite to the incident wave.

\begin{figure}
\centering
\includegraphics[width=10cm]{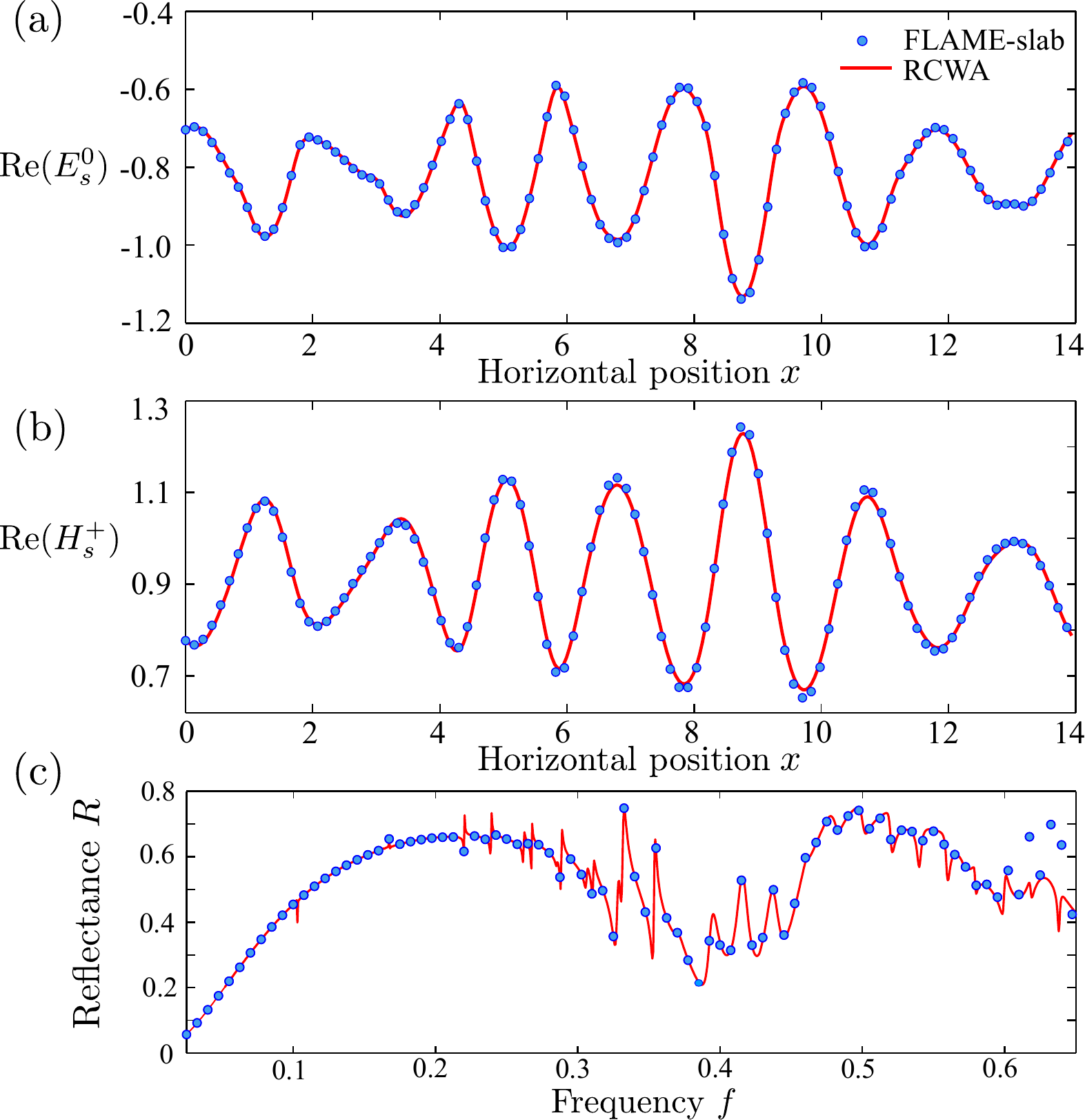}
\caption{(a) Real part of the electric field $E_{s}$ versus horizontal
  position $x$, along the middle layer of grid points at $z = z_0$.
  (b) Real part of the magnetic field $H_{s}$ versus $x$, along the
  outer layer of grid points at $z = z_-$.  In both cases, blue dots
  show FLAME-slab results calculated using $N_xw/L_x = 1.44$, $N_T
  w/L_1 = 160$, and $N_G = 150$; red curves show reference solutions
  calculated using RCWA with $N_G^{\mathrm{ref}} = 1000$.  The
  frequency is $f = 0.25$, and all other parameters are as stated in
  the main text.  (c) Reflectance $R$ versus frequency.  Blue dots
  show FLAME-slab results, obtained using $N_xw/L_x = 1.44$,
  $N_Tw/L_1=160$, and $N_G=97$; the red curve shows RCWA results
  calculated with $N_G^{\mathrm{ref}} = 1000$.  }
\label{fig:fields}
\end{figure}

Figure~\ref{fig:fields}(a)--(b) shows a typical set of electric and
magnetic fields calculated for the test structure.  Here, we plot
$\mathrm{Re}[E_s^0]$ and $\mathrm{Re}[H_s^-]$ versus $x$, comparing
the FLAME-slab results to a reference solution obtained by applying
RCWA to the entire structure.  For the FLAME-slab calculation, we take
$N_x w/L_x = 4.30$, $N_Tw/L_1 = 160$, and $N_G = 150$.  For the RCWA
reference solution, we take an ``overkill' value
$N_G^{\mathrm{ref}}=1000$ to ensure high accuracy for the big global
structure.  From the figure, we observe that the FLAME-slab solution
is in excellent agreement with the reference one.  The results for the
other field components show similar agreement.

Figure~\ref{fig:fields}(c) plots the reflectance $R$ versus frequency
$f$.  For frequencies $f \lesssim 0.5$, the FLAME-slab results are
again in close agreement with the reference solution.  The
transmittance is not shown here, but we verified that it satisfies the
flux conservation condition $R + T = 1$ to high precision.  At higher
frequencies, $f \gtrsim 0.5$, numerical errors become evident.  A
similar frequency-dependence of the numerical errors has also been
observed in the field components. It is to be expected that higher
frequencies require finer grids and other parameter adjustments.

\subsection{Error Analysis}
\label{sec:flame-error}

\begin{figure}
\centering
\includegraphics[width=8cm]{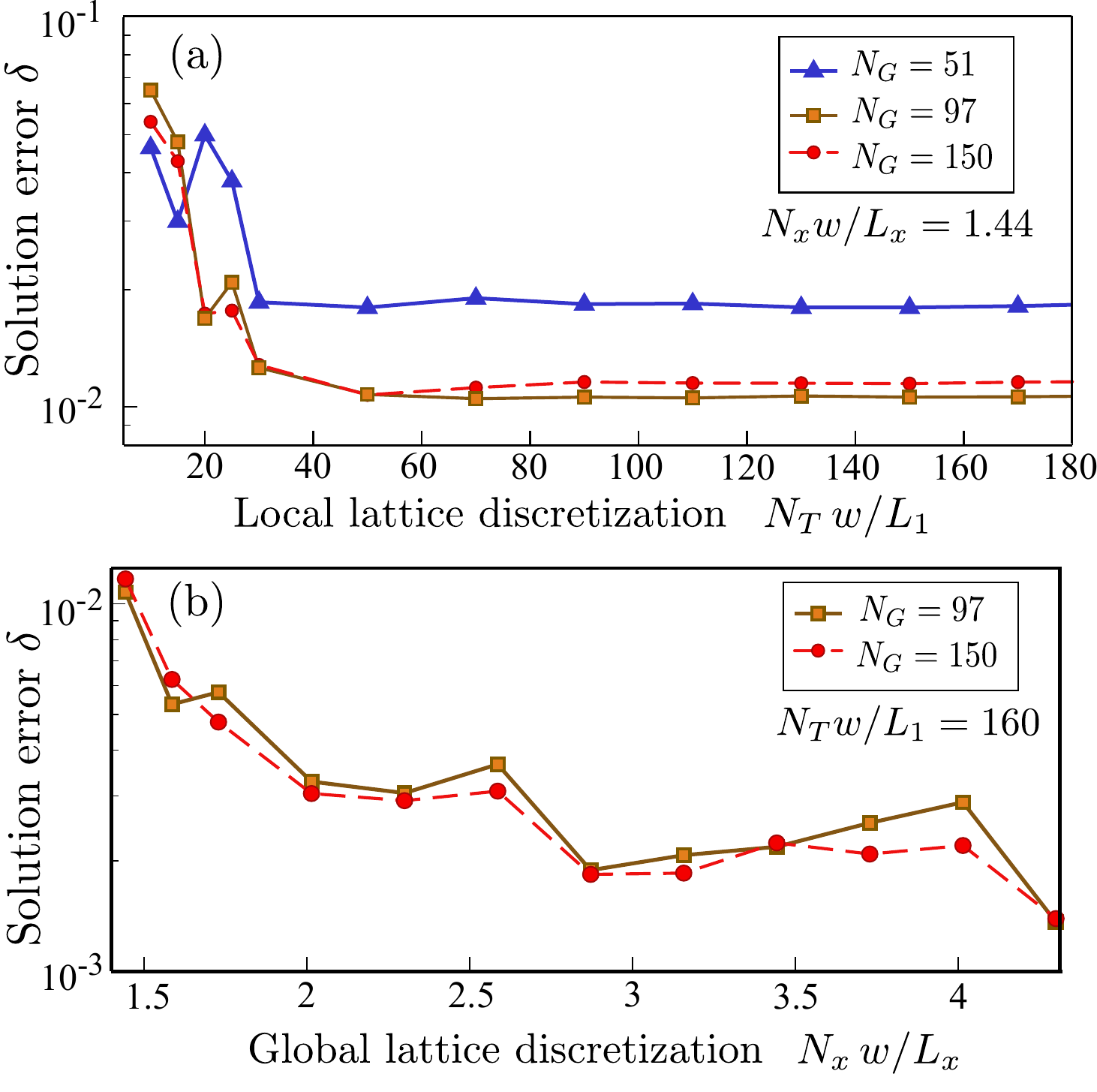}
\caption{(a) Solution error $\delta$ versus the discretization of the
  local lattice used for computing the Trefftz functions, $N_T w/L_1$.
  (b) Solution error $\delta$ versus the discretization of the global
  lattice, $N_x w / L_x$. The frequency is $f = 0.25$, and the
  reference solution is computed by RCWA using $N_G^{\mathrm{ref}} =
  1000$.}
\label{fig:solution_error}
\end{figure}

As mentioned in Section~\ref{sec:parameters}, there are three main
sources of error in FLAME-slab:
\begin{enumerate}
\item Even though FLAME schemes are qualitatively more accurate than
  the classical ones, FD-FLAME is still subject to discretization
  errors arising from the finite size of the Trefftz basis and
  DoF. This error depends on the dimensionless parameter $N_xw/L_x$.

\item The RCWA solutions are calculated on a grid, and interpolation
  is used to evaluate the Trefftz basis functions in each local
  stencil; this is controlled by $N_Tw/L_1$.

\item The basis functions are calculated using RCWA, which itself has
  finite accuracy; this is controlled by the paremeter $N_G$.
\end{enumerate}

To quantify how the accuracy of FLAME-slab depends on these
parameters, we perform a comparison to the reference solution noted
above.  We define a normalized ``solution error'',
\begin{equation}
  \delta =  \left(\frac{\displaystyle\sum_{i=1}^{M} \left| \psi_{i}^{\mathrm{Fs}}-\psi_{i}^{\mathrm{ref}} \right|^2}{\displaystyle \sum_{i=1}^{M} \left| \psi_{i}^{\mathrm{ref}} \right|^2}\right)^{1/2}.
  \label{eq:sol_error}
\end{equation}
Here, $\psi_{i}^{\mathrm{Fs}}$ denotes components of the FLAME-slab
solution vector (see Section~\ref{sec:slab}), consisting of the $E$
fields on all nodes and $H_x$ field components along the boundary
nodes; and $\psi_i^{\mathrm{ref}}$ denotes the corresponding field
components for the reference solution.  The sums in the numerator and
denominator are taken over all $M = 5N_x$ variables in the FLAME-slab
system of equations.

The specific choices of $N_x$, $N_T$, and $N_G$ determine which of the
three parameters dominate the solution error $\delta$ at any one time.
For instance, Fig.~\ref{fig:solution_error}(a) shows that for a fixed
value of $N_x$, $\delta$ decreases rapidly and then saturates.  This
saturated value decreases as we increase $N_G$ from 51 to 97, but
stops falling further as we increase $N_G$ to 150.  Next, suppose that
we fix $N_Tw/L_1 = 160$ (well within the region where $\delta$ is
saturated), and increase $N_x$.  As shown in
Fig.~\ref{fig:solution_error}(b), this causes $\delta$ to fall
rapidly, before saturating again.

\subsection{Runtime and memory performance}

\begin{figure}[b]
  \centering
  \includegraphics[width=11cm]{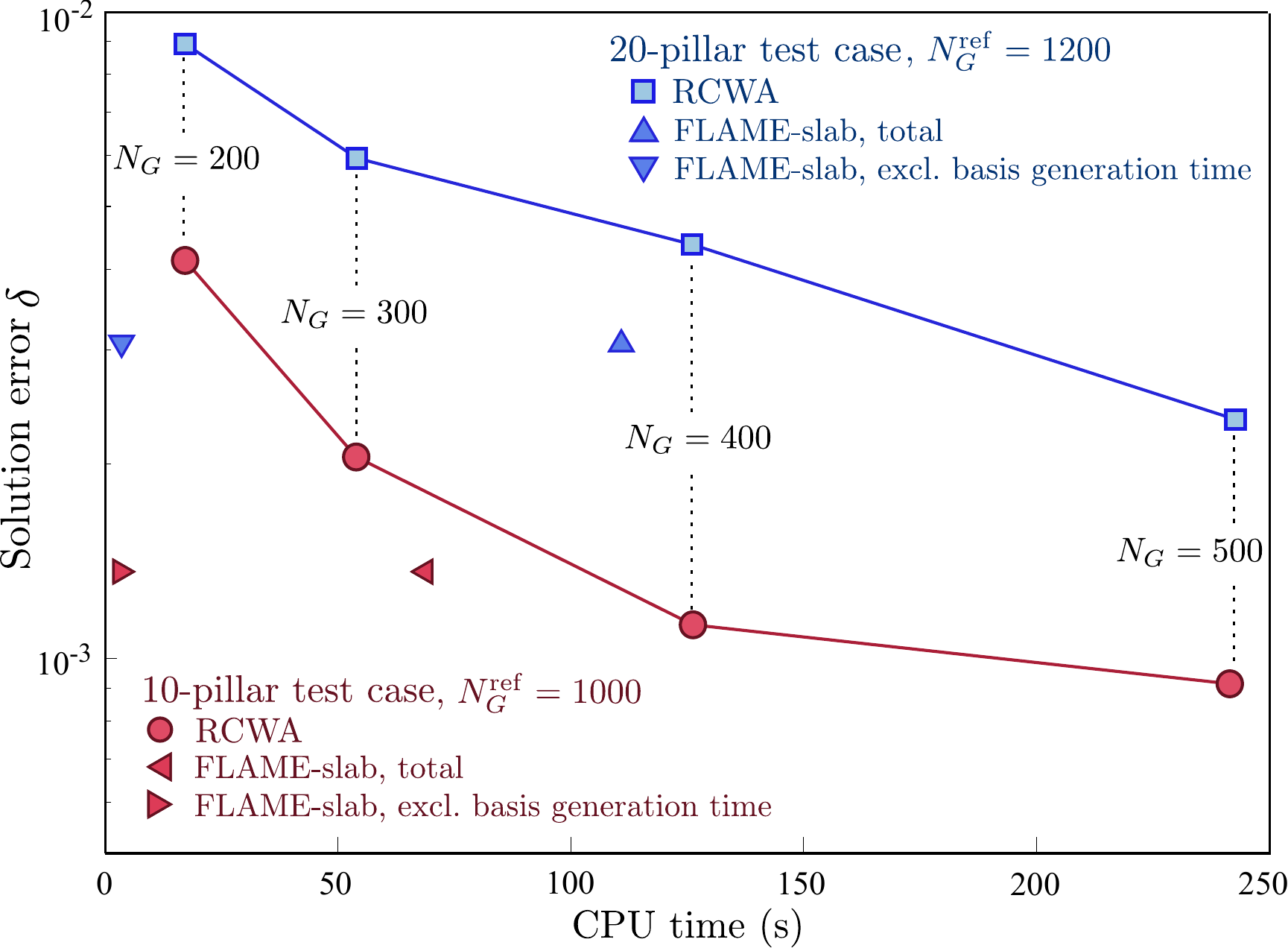}
  \caption{Solution error versus runtime for FLAME-slab and RCWA solutions, with two differently-sized test cases: (i) 10 pillars and $L_x = 14$, and (ii) 20 pillars and $L_x = 28$.  The FLAME-slab solutions were calculated with $N_xw/L_x = 4.30$ and $N_Tw/L_1 = 160$; and $N_G = 97$ for case (i), and $N_G = 120$ for case (ii).  The RCWA solutions were calculated with $N_G = 200, \, 300, \,400,\, 500$.  The reference solutions were calculated by RCWA using $N_G = 1000$ for case (i), and $N_G = 1200$ for case (ii).}
	\label{fig:runtime}
\end{figure}

We have profiled the runtime and memory performance for FLAME-slab calculations against calculations performed purely using RCWA.

Figure~\ref{fig:runtime} plots the solution error versus runtime for
FLAME-slab and RCWA calculations.  Two test cases are studied: (i) a
slab with 10 pillars and $L_x = 14$, identical to the structure tested
in Sections~\ref{sec:fields}--\ref{sec:flame-error}, and (ii) a larger
slab with 20 pillars and $L_x = 28$.  In both cases, the pillars are
placed randomly, without overlap, along the length of the slab.  For
the first case, the reference RCWA solution used to determine the
solution error was computed with $N_G^{\mathrm{ref}} = 1000$; for the
second case, the reference RCWA solution was computed with
$N_G^{\mathrm{ref}} = 1200$, the maximum possible before running out
of memory on the desktop computer performing these tests.

For RCWA, we observe that an increase in $N_G$ reduces the solution
error, while increasing the runtime (as $N_G^3$).  For FLAME-slab, the
\textit{total} runtime can be competitive with or superior to the
runtime of an RCWA calculation at a comparable level of solution
error, so long as the parameters $\{N_x, N_T, N_G\}$ are chosen
appropriately.  In Fig.~\ref{fig:runtime}, for example, the FLAME-slab
total runtime is shown to be $\approx 70\%$ of the RCWA runtime with
comparable solution error, for the 10-pillar case; for the 20-pillar
case, it is $\approx 60\%$.

FLAME-slab acquires a dramatic advantage if the Trefftz basis
functions can be saved and re-used.  Most of the total runtime of
FLAME-slab is spent on the initial calculation of the Trefftz basis
functions.  As we have previously noted, numerical studies of
aperiodic photonic structures are often concerned with
\textit{ensembles} of independent samples with the same regular
elements, such as amorphous or quasicrystalline lattices built out of
identical pillars.  With FLAME-slab, the Trefftz basis functions can
be computed just once and re-used; when we use RCWA or other
traditional methods, by comparison, the calculation must be re-done
entirely for each sample.

To illustrate the magnitude of this advantage, Fig.~\ref{fig:runtime}
shows also the FLAME-slab runtimes \textit{excluding} the time taken
to pre-compute the Trefftz basis functions.  These are around 2 orders
of magnitude smaller than RCWA runtimes with comparable solution
error.

This numerical data was obtained using a direct solver, which required
all matrix elements to be computed explicitly.  As discussed in
Section~\ref{sec:assembly}, it should be possible to achieve an
additional speedup by avoiding direct construction of the
$\mathbf{A}_{\mathrm{BC}}$ matrix, and instead using an iterative
solver with matrix-vector products evaluated by Fast Fourier
Transforms.

The memory consumption of FLAME-slab is significantly lower than that
of full-scale RCWA calculations.  For $N_G = 400$, the $S^4$ program
uses $198\,\mathrm{MB}$ of memory.  As seen in Fig.~\ref{fig:runtime},
FLAME-slab achieves a similar solution error with the parameters
$N_xw/L_x = 4.30$, $N_Tw/L_1 = 160$, and $N_G = 97$.  The memory usage
for FLAME-slab, with these parameters, consists of: (i)
pre-computation of the Trefftz functions using RCWA, which takes
$\approx 11.6\,\mathrm{MB}$ of memory; (ii) storage of the Trefftz
functions, which takes $\approx 1.5\,\mathrm{MB}$; (iii) storage of
the $\mathbf{A}$ matrix, which occupies $\approx 4.4\,\mathrm{MB}$ of
memory; and (iv) the memory usage of the direct solver, which we
estimated as $\approx 6.6\,\mathrm{MB}$ based on the sparsity of the
LU decomposition.  Thus, the overall memory usage of FLAME-slab in
this example is an order of magnitude lower than that of S$^4$.

\section{Conclusion}

We have devised a numerical method, FLAME-slab, which is suitable for
solving the wave scattering problem in photonic slabs lacking in-plane
periodicity.  FLAME \cite{Tsukerman2005} is used to generate
finite-difference stencils from a set of basis functions. In this
case, the basis functions are computed by running a subroutine---the
$S^4$ RCWA solver \cite{Liu2012}---on a set of local problems defined
on small-scale ``Trefftz cells''.

One advantage of FLAME-slab over traditional finite-difference and
finite-element methods is that a relatively coarse grid can be used
without sacrificing accuracy.  In the present implementation, the grid
consists of only three horizontal layers ($z_+$, $z_0$, and $z_-$).
Another key advantage is that for slab structures with short-range
regularities, such as pillars that are placed randomly but have
identical geometries and materials, the basis functions only have to
be evaluated once.  A further improvement, which we have not yet
extensively explored, is to use an iterative solver.

In future work, we seek to implement the scheme for fully 3D slab
structures.  We will also investigate how best to treat the outgoing
boundary conditions; it may be possible to use boundary element
techniques, perfectly-matched layers, or absorbing boundary conditions
instead of FFTs. It will also be interesting to explore methods other
than RCWA to generate the basis functions, including the Vertical Mode
Expansion Method \cite{Lu2014}.

\section{Funding}

This research was supported by the Singapore MOE Academic Research
Fund Tier 2 Grant MOE2016-T2-1-128, the Singapore MOE Academic
Research Fund Tier 2 Grant MOE2015-T2-2-008, and the Singapore MOE
Academic Research Fund Tier 3 Grant MOE2016-T3-1-006. The work of IT
was supported in part by the US National Science Foundation Grants
DMS-1216927 and DMS-1620112.

\end{document}